\documentclass[executivepaper,openbib,final,11pt,notitlepage]{article}
\usepackage{amsfonts}
\usepackage{amsmath}

\input{tcilatex}

\begin{document}

\title{New Solvable Shape-Invariant Potentials for Position-Dependent
Effective Mass}
\author{S.-A. Yahiaoui, H. Zerguini, M. Bentaiba\thanks{%
Corresponding author : bentaiba@hotmail.com.}. \\
LPTHIRM, D\'{e}partement de Physique, Facult\'{e} des Sciences,\\
Universit\'{e} Saad DAHLAB de Blida, Alg\'{e}rie.}
\maketitle

\begin{abstract}
Four new exactly solvable, real and shape-invariant potentials associated
with a position-dependent effective mass are generated within the concept of
shape-invariant potentials using a specific ansatz for superpotential. The
accompanying energy spectra of the bound-state and the ground-state
wavefunction are obtained algebraically as a function of free parameters and
the results are compared with those of others works in the litterature.



\ \ \ \ \ \ \ \ \ \ \ \ \ \ \ Shape-Invariant Potential.
\end{abstract}

\section{Introduction}

Physical systems with position-dependent effective mass have received, in
recent years, a significant attention due to their relevance in describing
the physical properties of various microstructures such as compositionally
graded crystals [1], quantum dots [2], semi conductor heterostructures [3],
quantum liquids [4] and $^{3}H-$clusters [5], etc. Recently, a wide number
of exact solutions on these topics has increased [6-11]. In theoretical
physics, various methods and approaches are used including Factorization and
Operator methods [12,13], Point canonical transformation methods [14],
Supersymmetric quantum mechanical approach [15], Group theory approach [16]
and Path integral formalism [17].

The effective mass Schr\"{o}dinger equation is studied by many authors and
its exact solutions (eigenvalues and eigenfunctions) are obtained. Quesne
and Tkachuk have established a certain connection between the Schr\"{o}%
dinger equation with position-dependent effective mass and the deforming
function appearing in the generalized canonical commutation relations. As a
consequence, the potential in the deformed Schr\"{o}dinger equation can be
considered as the effective potentials in position-dependent effective mass
[18,19].

The main purpose of the present paper is to obtain the energy spectra of the
bound-states and the ground-state wavefunctions for four \textquotedblright
new\textquotedblright\ solvable and real potentials (Three-dimensional
Harmonic Oscillator, Morse, P\"{o}schl-Teller I and II, and
three-dimensional Coulomb) deduced by applying the procedure of Ref. [18] in
order to solve a position-dependent effective mass as a deformed
shape-invariance condition introduced by Gendenshtein [20] and inspired by
the supersymmetric quantum mechanical technique [21]. The one-dimensional
potentials thus obtained are Shape-Invariant under parameter translations $%
\mathbf{\lambda }_{i+1}=\mathbf{\lambda }_{i}\pm \boldsymbol{a}$ where $%
\mathbf{\lambda }_{i}$ and $\boldsymbol{a}$\ denote two sets of three
parameters, i.e. $\mathbf{\lambda }_{i}=\left( \lambda _{i},\sigma _{i},\rho
_{i}\right) $\ and $\boldsymbol{a}=\left( a,b,c\right) $\ with $%
i=0,1,2,\ldots $ The new shape-invariance\ yields a new energy spectra of
the bound-states with a non-equidistant spectrum which present some features
with those obtained in Ref. [25].

The plan of the present paper is as follow. Factorization, effective
potential and Shape-Invariant Potentials are briefly reviewed in section 2.
Section 3 will deal with an assumption carried with the superpotential in
order to discuss the shape invariance condition leading to the class of
potentials and its corresponding \textquotedblright new\textquotedblright\
energy eigenvalues as well as their ground-state wavefunction. Finally, the
last section contain the conclusion.

\section{Factorization, Effective potential and Shape Invariant Potentials.}

There are several ways to define the kinetic energy operator when the mass
is a function of position. Since the mass and momentum operators are no
longer commute, the generalization of the standard Hamiltonian is not
trivial. Therefore, the choice of the correct ordering of the operators of
the kinetic energy to be Hermitian is indispensable. Defining a general
Hermitian effective mass Hamiltonian proposed by von Roos [3]%
\begin{equation}
\mathcal{H}_{VR}=\frac{1}{4}\left( m^{\alpha }\left( x\right) pm^{\beta
}\left( x\right) pm^{\gamma }\left( x\right) +m^{\gamma }\left( x\right)
pm^{\beta }\left( x\right) pm^{\alpha }\left( x\right) \right) +V\left(
x\right) ,  \tag{2.1}
\end{equation}%
where $\hbar =1,\ \alpha +\beta +\gamma =-1$, and $p=-i\dfrac{d}{dx}$. The
limits of the choice of the parameters $\alpha ,\beta $ and $\gamma $\
depend on the physical system. Using the restricted Hamiltonian from the $%
\alpha =\gamma =0$\ constraint used by BenDaniel and Duke [22], we can write
(2.1) as%
\begin{equation}
\mathcal{H}_{VR}=-\partial _{x}U^{2}\left( x\right) \partial _{x}+V\left(
x\right) ,  \tag{2.2}
\end{equation}%
with $U^{2}\left( x\right) =\dfrac{1}{2m\left( x\right) }$. Here we have
used abbreviation $\partial _{x}=\dfrac{d}{dx}.$

The identity found upon the commutation relation%
\begin{eqnarray}
\left[ \partial _{x},\sqrt{U\left( x\right) }\right]  &=&\partial _{x}\sqrt{%
U\left( x\right) }-\sqrt{U\left( x\right) }\partial _{x}  \notag \\
&=&\frac{1}{2}\frac{\partial _{x}U\left( x\right) }{\sqrt{U\left( x\right) }}%
,  \TCItag{2.3}
\end{eqnarray}%
brings the kinetic term in (2.2) to%
\begin{eqnarray}
\partial _{x}U^{2}\left( x\right) \partial _{x} &=&\partial _{x}\sqrt{%
U\left( x\right) }U\left( x\right) \sqrt{U\left( x\right) }\partial _{x} 
\notag \\
&=&\left[ \sqrt{U\left( x\right) }\partial _{x}+\frac{1}{2}\frac{\partial
_{x}U\left( x\right) }{\sqrt{U\left( x\right) }}\right] U\left( x\right) %
\left[ \partial _{x}\sqrt{U\left( x\right) }-\frac{1}{2}\frac{\partial
_{x}U\left( x\right) }{\sqrt{U\left( x\right) }}\right]   \notag \\
&=&\left( \sqrt{U\left( x\right) }\partial _{x}\sqrt{U\left( x\right) }%
\right) ^{2}-\frac{U^{\prime \prime }\left( x\right) U\left( x\right) }{2}-%
\frac{U^{\prime 2}\left( x\right) }{4}.  \TCItag{2.4}
\end{eqnarray}%
where the prime $"\ ^{\prime }\ "$ refers to the derivetive of $U\left(
x\right) $ with respect to $x$.

The Hamiltonian (2.1) becomes [18,19]%
\begin{equation}
\mathcal{H}_{VR}=-\left( \sqrt{U\left( x\right) }\partial _{x}\sqrt{U\left(
x\right) }\right) ^{2}+V_{eff}\left( x\right) ,  \tag{2.5}
\end{equation}%
where the effective potential $V_{eff}\left( x\right) $ is defined following
(2.1) and (2.4) as%
\begin{equation}
V_{eff}\left( x\right) =V\left( x\right) +\mathcal{V}_{U}\left( x\right) , 
\tag{2.6}
\end{equation}%
with%
\begin{equation}
\mathcal{V}_{U}\left( x\right) =\frac{U^{\prime \prime }\left( x\right)
U\left( x\right) }{2}+\frac{U^{\prime 2}\left( x\right) }{4}.  \tag{2.7}
\end{equation}

In the formalism of supersymmetric quantum mechanics, there are two
operators $Q$ and $Q^{\dag }$, called supercharges, that satisfy the anti
commutation relation $\left\{ Q,Q^{\dag }\right\} =\mathcal{H}_{SS}$, where $%
\mathcal{H}_{SS}$\ is the supersymmetric Hamiltonian [21]. The standard
realization of the operators $Q$ and $Q^{\dag }$ is $Q=A\sigma _{-}$ and $%
Q^{\dag }=A^{\dag }\sigma _{+}$\ where $A\left( A^{\dag }\right) $\ and $%
\sigma _{-}\left( \sigma _{+}\right) $\ are the bosonic operators and Pauli
matrices, respectively. As a consequence of this, the appropriate operators
to study the Hamiltonian (2.5) are%
\begin{eqnarray}
A_{eff} &=&\sqrt{U\left( x\right) }\partial _{x}\sqrt{U\left( x\right) }%
+W_{eff}\left( x\right) ,  \TCItag{2.8.a} \\
A_{eff}^{\dag } &=&-\sqrt{U\left( x\right) }\partial _{x}\sqrt{U\left(
x\right) }+W_{eff}\left( x\right) .  \TCItag{2.8.b}
\end{eqnarray}%
where $W_{eff}\left( x\right) $\ is called the effective superpotential.

With this realization, the supersymmetric Hamiltonian (2.5) of the quantum
system with position-dependent effective mass takes the form%
\begin{eqnarray}
\mathcal{H}_{1,eff} &\equiv &A_{eff}^{\dag }A_{eff}  \notag \\
&=&-\left( \sqrt{U\left( x\right) }\partial _{x}\sqrt{U\left( x\right) }%
\right) ^{2}+V_{1,eff}\left( x\right) ,  \TCItag{2.9.a}
\end{eqnarray}%
where%
\begin{equation}
V_{1,eff}\left( x\right) =W_{eff}^{2}\left( x\right) -U\left( x\right)
W_{eff}^{\prime }\left( x\right) ,  \tag{2.9.b}
\end{equation}%
and%
\begin{eqnarray}
\mathcal{H}_{2,eff} &\equiv &A_{eff}A_{eff}^{\dag }  \notag \\
&=&-\left( \sqrt{U\left( x\right) }\partial _{x}\sqrt{U\left( x\right) }%
\right) ^{2}+V_{2,eff}\left( x\right) ,  \TCItag{2.10.a}
\end{eqnarray}%
with%
\begin{equation}
V_{2,eff}\left( x\right) =W_{eff}^{2}\left( x\right) +U\left( x\right)
W_{eff}^{\prime }\left( x\right) .  \tag{2.10.b}
\end{equation}

The Hamiltonian $\mathcal{H}_{2,eff}$\ is called the supersymmetric partner
of $\mathcal{H}_{1,eff}$\ . It can be easily shown that both $\mathcal{H}%
_{1,eff}$\ and $\mathcal{H}_{2,eff}$\ has the same spectrum except for the
ground-state, which belongs to $\mathcal{H}_{1,eff}$ [21].\ It is obvious
that (2.9.b) and (2.10.b) are related by%
\begin{equation}
V_{2,eff}\left( x\right) =V_{1,eff}\left( x\right) +2U\left( x\right)
W_{eff}^{\prime }\left( x\right) .  \tag{2.11}
\end{equation}

Substituting $\mathcal{V}_{U}\left( x\right) $ given by (2.7) as defined in
(2.6) into $V_{i,eff}\left( x\right) $ with $i=1,2$\ given by (2.11), we end
up with%
\begin{equation}
V_{2}\left( x\right) =V_{1}\left( x\right) +2U\left( x\right)
W_{eff}^{\prime }\left( x\right) .  \tag{2.12}
\end{equation}

However, following the paper of Samani and Loran [23], the potential $%
V_{2}\left( x\right) $ reads as%
\begin{equation}
V_{2}\left( x\right) =V_{1}\left( x\right) +2U\left( x\right) W^{\prime
}\left( x\right) -U\left( x\right) U^{\prime \prime }\left( x\right) , 
\tag{2.13}
\end{equation}%
and comparing (2.12) to (2.13), we obtain after integration the relationship
connecting both the superpotential and the effective superpotential%
\begin{equation}
W\left( x\right) =W_{eff}\left( x\right) +\frac{U^{\prime }\left( x\right) }{%
2}.  \tag{2.14}
\end{equation}

Inserting (2.14) into (2.9.b) and (2.10.b), we get [23]%
\begin{eqnarray}
V_{1}\left( x\right) &=&W^{2}\left( x\right) -\left[ U\left( x\right)
W\left( x\right) \right] ^{\prime },  \TCItag{2.15.a} \\
V_{2}\left( x\right) &=&W^{2}\left( x\right) -\left[ U\left( x\right)
W\left( x\right) \right] ^{\prime }+2U\left( x\right) W^{\prime }\left(
x\right)  \notag \\
&&-U\left( x\right) U^{\prime \prime }\left( x\right) .  \TCItag{2.15.b}
\end{eqnarray}

Despite their similar bound-state energy spectra, supersymmetric partner
potentials constructed from (2.9-10.b) and (2.15.a-b) usually have different
structures. However, the above potentials are called shape-invariant if $%
V_{2}\left( x\right) $\ has the same functional dependence on the coordinate
as $V_{1}\left( x\right) $\ and differ only in some parameters [21]. The
shape-invariant potentials are defined by the relationship%
\begin{eqnarray}
V_{2}\left( x,a_{0}\right) -V_{1}\left( x,a_{1}\right)  &=&2U\left( x\right)
W^{\prime }\left( x,a_{0}\right) -U\left( x\right) U^{\prime \prime }\left(
x\right)   \notag \\
&=&\mathcal{R}\left( a_{0}\right) ,  \TCItag{2.16}
\end{eqnarray}%
where $a_{1}=y\left( a_{0}\right) $\ is a function of parameters $a_{0}$\
and $\mathcal{R}\left( a_{0}\right) $\ is independent of variable $x.$\ As a
consequence, it can be shown that the discrete energy spectrum of $%
V_{1}\left( x\right) $\ can be written as [21]

\begin{equation}
E_{n}=\dsum\limits_{k=0}^{n-1}\mathcal{R}\left( a_{k}\right) ,  \tag{2.17}
\end{equation}%
where the parameter $a_{k}$\ is generated by the consecutive application of
the function $y\left( x\right) ,$ i.e.%
\begin{equation}
a_{k}\equiv y^{\left( k\right) }\left( a_{0}\right) =\underset{\text{k-times}%
}{\underbrace{y\circ y\circ \cdots \circ y\left( a_{0}\right) }}.  \tag{2.18}
\end{equation}

Having determined the bound-state energy spectrum, the ground-state
wavefunction of the corresponding Hamiltonian is obtainable by solving the
first-order differential equation%
\begin{equation}
A_{eff}\psi _{0}\left( x\right) =0,  \tag{2.19}
\end{equation}%
leading to the wavefunction%
\begin{equation}
\psi _{0}\left( x\right) =\frac{\mathcal{N}_{0}}{\sqrt{U\left( x\right) }}%
\exp \left[ -\dint\limits^{z}dz\frac{W\left( z\right) }{U\left( z\right) }%
\right] ,  \tag{2.20}
\end{equation}%
where\ $\mathcal{N}_{0}$\ is some normalization coefficient.

\section{Exactly solvable potentials with three parameters}

The fairly general factorizable form of superpotential $W\left( x\right) $\
is given by Ref [21]%
\begin{equation}
W\left( x,a_{0}\right) =\dsum\limits_{i=1}^{s}\left( k_{i}+c_{i}\right)
g_{i}\left( x\right) +\frac{h_{i}\left( x\right) }{k_{i}+c_{i}}+f_{i}\left(
x\right) ,  \tag{3.1}
\end{equation}%
where $a_{0}=\left( k_{1},k_{2},\ldots \right) $, $a_{1}=\left( k_{1}+\alpha
,k_{2}+\beta ,\ldots \right) $ and $c_{i},\alpha ,\beta ,\ldots $ being
constants. It may be noted that one can obtain the condition to be fulfilled
by the functions $g_{i}\left( x\right) ,h_{i}\left( x\right) $ and $%
f_{i}\left( x\right) $ once that (3.1)\ is used in (2.16). However, Samani
and Loran have carried out a simple form to $W$\ considering it in the
ansatz with one-parameter [23]%
\begin{equation}
W\left( x,a\right) =ag\left( x\right) +\frac{h\left( x\right) }{a}+f\left(
x\right) ,  \tag{3.2}
\end{equation}%
where the functions $g\left( x\right) ,h\left( x\right) $ and $f\left(
x\right) $ are independent of the parameter $a.$

We shall now point out the specific ansatz carried out by the superpotential
that goes into the determination of the effective potentials and its
accompanying energy eigenvalues and ground-state wavefunctions. To this end,
the \textquotedblright new\textquotedblright\ superpotential is
characterized by one function and three parameters instead of that given by
(3.2). To be more precise, we substitute all functions appearing in (3.2) by
new parameters and the single parameter by a function, i.e. $g\left(
x\right) \rightarrow \lambda ,\ f\left( x\right) \rightarrow \sigma ,\
h\left( x\right) \rightarrow \rho $ and $a\rightarrow \phi \left( x\right) ,$%
\ without making a point of seeking the transformation which connects them.

Therefore the superpotential (3.2) becomes, taking into account (2.14)%
\begin{equation}
W\left( x,\mathbf{\lambda }\right) =\lambda \phi \left( x\right) +\frac{\rho 
}{\phi \left( x\right) }+\sigma +\frac{U^{\prime }\left( x\right) }{2}, 
\tag{3.3}
\end{equation}%
and the effective shape-invariance condition reads as%
\begin{equation}
V_{2,eff}\left( x,\mathbf{\lambda }_{0}\right) =V_{1,eff}\left( x,\mathbf{%
\lambda }_{1}\right) +\mathcal{R}\left( a_{0}\right) ,  \tag{3.4}
\end{equation}%
with $\mathbf{\lambda }_{i}=\left( \lambda _{i},\sigma _{i},\rho _{i}\right) 
$,$\ i=0,1,2,\ldots $ Henceforth, we will suppose that $\mathbf{\lambda }$
coincides with $\mathbf{\lambda }_{0}$.

Inserting (3.3) into (3.4) we obtain%
\begin{multline}
\left( \lambda _{1}^{2}-\lambda _{0}^{2}\right) \phi ^{2}\left( x\right)
+2\left( \lambda _{1}\sigma _{1}-\lambda _{0}\sigma _{0}\right) \phi \left(
x\right) +\frac{\rho _{1}^{2}-\rho _{0}^{2}}{\phi ^{2}\left( x\right) }+%
\frac{2\left( \rho _{1}\sigma _{1}-\rho _{0}\sigma _{0}\right) }{\phi \left(
x\right) }+  \notag \\
U\left( x\right) \left( \rho _{1}+\pi \rho _{0}\right) \frac{\phi ^{\prime
}\left( x\right) }{\phi ^{2}\left( x\right) }-U\left( x\right) \left(
\lambda _{1}+\lambda _{0}\right) \phi ^{\prime }\left( x\right)   \notag \\
+\sigma _{1}^{2}-\sigma _{0}^{2}+2\left( \rho _{1}\lambda _{1}-\rho
_{0}\lambda _{0}\right) +\mathcal{R}\left( \mathbf{\lambda }_{0}\right) =0. 
\tag{3.5}
\end{multline}%
with $\pi =\pm 1.$ The positive case $\left( \pi =1\right) $ is what one
ends up obtaining in (3.5) leading to the well-known results listed and
tabulated in Refs. [18,19,21], while the negative case $\left( \pi
=-1\right) $ was added thinking that it leads to more interesting one. In
the subsequent developments, we will be interested in the last case.

Then, one way for (3.5) to be consistent and solvable is to separate the
constant terms from the functions ; i.e.

\begin{multline}
\left( \lambda _{1}^{2}-\lambda _{0}^{2}\right) \phi ^{2}\left( x\right)
+2\left( \lambda _{1}\sigma _{1}-\lambda _{0}\sigma _{0}\right) \phi \left(
x\right) +\frac{\rho _{1}^{2}-\rho _{0}^{2}}{\phi ^{2}\left( x\right) }+%
\frac{2\left( \rho _{1}\sigma _{1}-\rho _{0}\sigma _{0}\right) }{\phi \left(
x\right) }+q  \notag \\
=U\left( x\right) \left( \lambda _{1}+\lambda _{0}\right) \phi ^{\prime
}\left( x\right) -U\left( x\right) \left( \rho _{1}-\rho _{0}\right) \frac{%
\phi ^{\prime }\left( x\right) }{\phi ^{2}\left( x\right) },  \tag{3.6}
\end{multline}%
and%
\begin{equation}
\sigma _{1}^{2}-\sigma _{0}^{2}+2\left( \rho _{1}\lambda _{1}-\rho
_{0}\lambda _{0}\right) +\mathcal{R}\left( \mathbf{\lambda }_{0}\right) =q, 
\tag{3.7}
\end{equation}%
with $q\neq 0.$ The resolution of (3.6) in terms of $\phi \left( x\right) $\
amounts to comparing these two members. Consequently, it is of primary
importance to write a term $U\left( x\right) \phi ^{\prime }\left( x\right) $
in such a way that the member of left-hand side is identified to that of
right-hand side. To this end, three particular cases (solutions) arise, and
henceforth, we will name them constant, linear\textit{\ }and\textit{\ }%
quadratic solutions and are identified with differential equations,
respectively%
\begin{eqnarray}
U\left( x\right) \phi ^{\prime }\left( x\right)  &=&a,  \TCItag{3.8.a} \\
U\left( x\right) \phi ^{\prime }\left( x\right)  &=&a\phi \left( x\right)
+b,\quad \left( a<0\right) ,  \TCItag{3.8.b} \\
U\left( x\right) \phi ^{\prime }\left( x\right)  &=&a\phi ^{2}\left(
x\right) +b\phi \left( x\right) +c.  \TCItag{3.8.c}
\end{eqnarray}%
where $a,b$\ and $c$\ are constants. It is obvious that the solutions in $%
\phi \left( x\right) $\ can be explicitly carried out by Euler's type
integration [24].

\subsection{Constant solution : Three-dimensional harmonic oscillator}

\subsubsection{Superpotential and Effective potential}

After a simple integration, the function $\phi \left( x\right) $\ is given by%
\begin{equation}
\phi \left( x\right) =a\mu \left( x\right) +b,  \tag{3.9}
\end{equation}%
where $\left( a,b\right) \in 
\mathbb{R}
^{2}$ and $\mu \left( x\right) $\ is a function defined as a dimensionless
integral%
\begin{equation}
\mu \left( x\right) =\dint\limits^{z}\frac{dz}{U\left( z\right) }, 
\tag{3.10}
\end{equation}%
and which will appear frequently in subsequent subsections.

Using both (3.3) and (2.15.a) we obtain, respectively, the superpotential
and the corresponding effective potential%
\begin{equation}
W\left( x,\mathbf{\lambda }\right) =\lambda \left( a\mu \left( x\right)
+b\right) +\frac{\rho }{a\mu \left( x\right) +b}+\sigma +\frac{U^{\prime
}\left( x\right) }{2},  \tag{3.11}
\end{equation}%
\begin{eqnarray}
V_{eff}\left( x,\mathbf{\lambda }\right)  &=&\lambda ^{2}\left( a\mu \left(
x\right) +b\right) ^{2}+\frac{\rho \left( \rho +a\right) }{\left( a\mu
\left( x\right) +b\right) ^{2}}+2\lambda \sigma \left( a\mu \left( x\right)
+b\right)   \notag \\
&&+\frac{2\sigma \rho }{a\mu \left( x\right) +b}+\sigma ^{2}+2\lambda \rho
-a\lambda .  \TCItag{3.12}
\end{eqnarray}

It is obvious that if the $\sigma =b=0$\ constraint holds, then the
effective potential (3.12) is related to shape-invariant three-dimensional
harmonic oscillator potential%
\begin{equation}
V_{eff}^{\left( \text{H.O}\right) }\left( x,\mathbf{\lambda }\right)
=\lambda ^{2}a^{2}\mu ^{2}\left( x\right) +\frac{\rho \left( \rho +a\right) 
}{a^{2}\mu ^{2}\left( x\right) }+2\lambda \rho -a\lambda .  \tag{3.13}
\end{equation}

\subsubsection{Energy eigenvalues}

The energy eigenvalues can be calculated algebraically from (3.7). Indeed,
inserting (3.8.a) into (3.6) we obtain%
\begin{eqnarray}
\lambda _{1}^{2}-\lambda _{0}^{2} &=&\lambda _{1}\sigma _{1}-\lambda
_{0}\sigma _{0}=\rho _{1}\sigma _{1}-\rho _{0}\sigma _{0}=0,  \TCItag{3.14.a}
\\
\rho _{1}^{2}-\rho _{0}^{2} &=&-a\left( \rho _{1}-\rho _{0}\right) , 
\TCItag{3.14.b} \\
q &=&a\left( \lambda _{1}+\lambda _{0}\right) \neq 0.  \TCItag{3.14.c}
\end{eqnarray}

Solving (3.14) gets $\rho _{1}=-\left( \rho _{0}+a\right) ,\ \lambda
_{1}=\lambda _{0}$ and $\sigma _{1}=\sigma _{0}$, which satisfies the
recursion relations%
\begin{equation}
\rho _{k}=\left( -1\right) ^{k}\rho _{0}-\frac{a}{2}\left( 1-\left(
-1\right) ^{k}\right) ,\quad \lambda _{k}=\lambda _{0},\quad \sigma
_{k}=\sigma _{0},  \tag{3.15}
\end{equation}%
where $k=0,1,2,\ldots $\ Thus the energy eigenvalues are given, taking into
consideration (2.17)

\begin{eqnarray}
\mathcal{E}_{n} &=&\dsum\limits_{k=0}^{n-1}\mathcal{R}\left( \lambda
_{k},\sigma _{k},\rho _{k}\right)  \notag \\
&=&\dsum\limits_{k=0}^{n-1}4\lambda _{k}\left( \rho _{k}+a\right)  \notag \\
&=&4a\dsum\limits_{k=0}^{n-1}\lambda _{0}+4\dsum\limits_{k=0}^{n-1}\lambda
_{0}\left[ \left( -1\right) ^{k}\rho _{0}-\frac{a}{2}\left( 1-\left(
-1\right) ^{k}\right) \right]  \notag \\
&=&2a\lambda _{0}n+\lambda _{0}\left( a+2\rho _{0}\right) \left( 1-\left(
-1\right) ^{n}\right) .  \TCItag{3.16}
\end{eqnarray}

This spectrum presents some features with that obtained in formula (10) in
Ref. [25]. Hence the negative case is allowed to generate new
shape-invariant potentials with richer new energy spectra of the
bound-states with a non-equidistant spectrum.

\subsubsection{Ground-state wavefunction}

Since a particle is constrained to move in three-dimensional Harmonic
Oscillator, we set $\sigma _{k}=b=0\ \left( k=0,1,2,\ldots \right) ,\
\lambda _{0}=\dfrac{\omega }{2a}$\ and $\rho _{0}=al\ \left( l<0\right) .$\
Then, keeping in mind (3.10), the ground-state wavefunction is calculated
straightforwardly using (2.19), we finally get%
\begin{eqnarray}
\psi _{0}\left( x\right)  &=&\frac{\mathcal{N}_{0}}{\sqrt{U\left( x\right) }}%
\exp \left[ -\lambda _{0}a\dint\limits^{z}dz\frac{\mu \left( z\right) }{%
U\left( z\right) }-\frac{\rho _{0}}{a}\dint\limits^{z}\frac{dz}{U\left(
z\right) \mu \left( z\right) }-\frac{1}{2}\dint\limits^{z}\frac{dU\left(
z\right) }{U\left( z\right) }\right]   \notag \\
&=&\frac{\mathcal{N}_{0}}{U\left( x\right) }\mu ^{-l}\left( x\right) \exp %
\left[ -\frac{\omega }{4}\mu ^{2}\left( x\right) \right] .  \TCItag{3.17}
\end{eqnarray}

\subsection{Linear solution : Morse potential}

\subsubsection{Superpotential and Effective potential}

In this case, the function $\phi \left( x\right) $\ is given by%
\begin{equation}
\phi \left( x\right) =\frac{1}{a}\left( b-\exp \left[ -a\mu \left( x\right) %
\right] \right) .  \tag{3.18}
\end{equation}%
with $a\neq 0$. The corresponding superpotential and the effective potential
read as%
\begin{equation}
W\left( x,\mathbf{\lambda }\right) =\frac{\lambda }{a}\left( b-\exp \left[
-a\mu \left( x\right) \right] \right) +\frac{a\rho }{b-\exp \left[ -a\mu
\left( x\right) \right] }+\sigma +\frac{U^{\prime }\left( x\right) }{2}, 
\tag{3.19}
\end{equation}%
\begin{eqnarray}
V_{eff}\left( x,\mathbf{\lambda }\right)  &=&\frac{\lambda ^{2}}{a^{2}}%
\left( b-\exp \left[ -a\mu \left( x\right) \right] \right) ^{2}+\frac{%
a^{2}\rho \left( \rho +b\right) }{\left( b-\exp \left[ -a\mu \left( x\right) %
\right] \right) ^{2}}  \notag \\
&&+\frac{a\rho \left( 2\sigma -a\right) }{b-\exp \left[ -a\mu \left(
x\right) \right] }+\lambda \left( \frac{2\sigma }{a}+1\right) \left( b-\exp %
\left[ -a\mu \left( x\right) \right] \right)   \notag \\
&&+\sigma ^{2}+2\lambda \rho -b\lambda .  \TCItag{3.20}
\end{eqnarray}

From (3.20), if the $\rho =b=0$\ constraint holds then the effective
potential is related to shape-invariant Morse potential%
\begin{equation}
V_{eff}^{\left( \text{Morse}\right) }\left( x,\mathbf{\lambda }\right) =%
\frac{\lambda ^{2}}{a^{2}}\func{e}^{-2a\mu \left( x\right) }-\lambda \left( 
\frac{2\sigma }{a}+1\right) \func{e}^{-a\mu \left( x\right) }+\sigma ^{2}. 
\tag{3.21}
\end{equation}

\subsubsection{Energy eigenvalues}

Inserting (3.8.b) into (3.6), we get the system of parametric equations
according to%
\begin{eqnarray}
\lambda _{1}^{2}-\lambda _{0}^{2} &=&0,  \TCItag{3.22.a} \\
2\left( \lambda _{1}\sigma _{1}-\lambda _{0}\sigma _{0}\right) &=&a\left(
\lambda _{1}+\lambda _{0}\right) ,  \TCItag{3.22.b} \\
\rho _{1}^{2}-\rho _{0}^{2} &=&-b\left( \rho _{1}-\rho _{0}\right) , 
\TCItag{3.22.c} \\
2\left( \rho _{1}\sigma _{1}-\rho _{0}\sigma _{0}\right) &=&-a\left( \rho
_{1}-\rho _{0}\right) ,  \TCItag{3.22.d} \\
q &=&b\left( \lambda _{1}+\lambda _{0}\right) \neq 0.  \TCItag{3.22.e}
\end{eqnarray}%
and its solution lead to the following recursion relations%
\begin{equation}
\rho _{k}=\left( -1\right) ^{k}\rho _{0}-\frac{b}{2}\left( 1-\left(
-1\right) ^{k}\right) ,\quad \lambda _{k}=\lambda _{0},\quad \sigma
_{k}=\sigma _{0}+ak.  \tag{3.23}
\end{equation}

The energy eigenvalues can be calculated easily%
\begin{eqnarray}
\mathcal{E}_{n} &=&\dsum\limits_{k=0}^{n-1}\mathcal{R}\left( \lambda
_{k},\sigma _{k},\rho _{k}\right)  \notag \\
&=&\dsum\limits_{k=0}^{n-1}\left( 4b\lambda _{k}-2a\sigma _{k}+4b\rho
_{k}-a^{2}\right)  \notag \\
&=&n\left( 2b\lambda _{0}-2a\sigma _{0}-a^{2}n\right) +\lambda _{0}\left(
2\rho _{0}+b\right) \left( 1-\left( -1\right) ^{n}\right)  \TCItag{3.24}
\end{eqnarray}

Again, we obtain a richer spectra for the Morse potential as was done in
(3.16) for the three-dimensional harmonic oscillator. The energy eigenvalues
corresponding to the effective potential (3.21) can be obtained by setting
the $\rho _{0}=b=0$\ constraint%
\begin{equation}
\mathcal{E}_{n}^{\left( \text{Morse}\right) }=-a^{2}\left( \frac{\sigma _{0}%
}{a}+n\right) ^{2}+\sigma _{0}^{2}.  \tag{3.25}
\end{equation}

\subsubsection{Ground-state wavefunction}

The ground-state wavefunction $\psi _{0}\left( x\right) $\ associated to the
energy eigenvalues (3.21) is given after integration%
\begin{eqnarray}
\psi _{0}\left( x\right) &=&\frac{\mathcal{N}_{0}}{\sqrt{U\left( x\right) }}%
\exp \left[ \frac{\lambda _{0}}{a}\dint\limits^{z}dz\frac{\func{e}^{-a\mu
\left( z\right) }}{U\left( z\right) }-\frac{a}{2}\dint\limits^{z}\frac{dz}{%
U\left( z\right) }-\frac{1}{2}\dint\limits^{z}\frac{dU\left( z\right) }{%
U\left( z\right) }\right]  \notag \\
&=&\frac{\mathcal{N}_{0}}{U\left( x\right) }\func{e}^{-a\mu \left( x\right)
/2}\exp \left[ -\frac{\lambda _{0}}{a^{2}}\func{e}^{-a\mu \left( z\right) }%
\right] .  \TCItag{3.26}
\end{eqnarray}

\subsection{Quadratic solution : P\"{o}schl-Teller (I,\ II) and three
dimensional Coulomb potentials}

Integrating (3.8.c) in terms of the function $\phi \left( x\right) $,\ it is
straightforward to obtain (see 2.172 of Ref. [24])%
\begin{eqnarray}
\dint\limits^{\phi \left( x\right) }\frac{d\xi }{a\xi ^{2}+b\xi +c} &=&\frac{%
2}{\sqrt{\Delta }}\arctan \frac{b+2a\phi \left( x\right) }{\sqrt{\Delta }}%
\quad ,\quad \ \Delta >0  \TCItag{3.27.a} \\
&=&\frac{-2}{\sqrt{-\Delta }}\func{arctanh}\frac{b+2a\phi \left( x\right) }{%
\sqrt{-\Delta }},\quad \Delta <0  \TCItag{3.27.b} \\
&=&\frac{-2}{b+2a\phi \left( x\right) }\qquad \qquad \qquad ,\quad \ \Delta
=0  \TCItag{3.27.c}
\end{eqnarray}%
where $\Delta =-b^{2}+4ac.$ It is obvious that the integral appearing up is
equal to a dimensionless mass integral, i.e. $\mu \left( x\right) $.

We will see in subsequent developments that the first two cases are related
to trigonometric and hyperbolic P\"{o}schl-Teller potentials, respectively,
while the last case is associated to the three-dimensional Coulomb potential.

\subsubsection{P\"{o}schl-Teller I and II}

\paragraph{Superpotential and Effective potential}

The solution in terms of $\phi \left( x\right) $\ is given through (3.27.a)
by $\phi \left( x\right) =\dfrac{\sqrt{\Delta }}{2a}\tan \dfrac{\sqrt{\Delta 
}}{2}\mu \left( x\right) -\dfrac{b}{2a},$\ the superpotential and the
effective potential read as%
\begin{equation}
W\left( x,\mathbf{\lambda }\right) =\lambda \left( \frac{\sqrt{\Delta }}{2a}%
\tan \frac{\sqrt{\Delta }}{2}\mu \left( x\right) -\frac{b}{2a}\right) +\frac{%
\rho }{\frac{\sqrt{\Delta }}{2a}\tan \frac{\sqrt{\Delta }}{2}\mu \left(
x\right) -\frac{b}{2a}}+\sigma +\frac{U^{\prime }\left( x\right) }{2}. 
\tag{3.28}
\end{equation}%
\begin{eqnarray}
V_{eff}\left( x,\mathbf{\lambda }\right)  &=&\lambda ^{2}\left( \frac{\sqrt{%
\Delta }}{2a}\tan \frac{\sqrt{\Delta }}{2}\mu \left( x\right) -\frac{b}{2a}%
\right) ^{2}+\frac{\rho ^{2}}{\left( \frac{\sqrt{\Delta }}{2a}\tan \frac{%
\sqrt{\Delta }}{2}\mu \left( x\right) -\frac{b}{2a}\right) ^{2}}  \notag \\
&&+2\lambda \sigma \left( \frac{\sqrt{\Delta }}{2a}\tan \frac{\sqrt{\Delta }%
}{2}\mu \left( x\right) -\frac{b}{2a}\right) +\frac{2\rho \sigma }{\frac{%
\sqrt{\Delta }}{2a}\tan \frac{\sqrt{\Delta }}{2}\mu \left( x\right) -\frac{b%
}{2a}}  \notag \\
&&-\frac{\frac{\lambda \Delta }{4a}}{\cos ^{2}\frac{\sqrt{\Delta }}{2}\mu
\left( x\right) }+\frac{\frac{\rho \Delta }{4a}}{\cos ^{2}\frac{\sqrt{\Delta 
}}{2}\mu \left( x\right) \left( \frac{\sqrt{\Delta }}{2a}\tan \frac{\sqrt{%
\Delta }}{2}\mu \left( x\right) -\frac{b}{2a}\right) ^{2}}  \notag \\
&&+\sigma ^{2}+2\lambda \rho .  \TCItag{3.29}
\end{eqnarray}

The $\sigma =b=0$\ constraint leads to the shape-invariant trigonometric P%
\"{o}schl-Teller potential. Indeed, taking into account $\Delta =4ac>0$, the
effective potential is reduced to%
\begin{equation}
V_{eff}^{\left( \text{Trig.}\right) }\left( x\right) =\frac{\lambda c\left( 
\frac{\lambda }{a}-1\right) }{\cos ^{2}\sqrt{ac}\mu \left( x\right) }+\frac{%
\rho a\left( \frac{\rho }{c}+1\right) }{\sin ^{2}\sqrt{ac}\mu \left(
x\right) }-\left( \lambda \sqrt{\frac{c}{a}}-\rho \sqrt{\frac{a}{c}}\right)
^{2}.  \tag{3.30}
\end{equation}

The hyperbolic P\"{o}schl-Teller superpotential and effective potential are
obtainable once the substitution $\sqrt{\Delta }\rightarrow i\sqrt{-\Delta }$%
\ is made. As a consequence of this, the superpotential and the effective
potential become respectively%
\begin{eqnarray}
W\left( x,\mathbf{\lambda }\right)  &=&-\lambda \left( \frac{\sqrt{-\Delta }%
}{2a}\tanh \frac{\sqrt{-\Delta }}{2}\mu \left( x\right) +\frac{b}{2a}\right)
-\frac{\rho }{\frac{\sqrt{-\Delta }}{2a}\tanh \frac{\sqrt{-\Delta }}{2}\mu
\left( x\right) +\frac{b}{2a}}  \notag \\
&&+\sigma +\frac{U^{\prime }\left( x\right) }{2}.  \TCItag{3.31}
\end{eqnarray}%
\begin{eqnarray}
V_{eff}\left( x,\mathbf{\lambda }\right)  &=&\lambda ^{2}\left( \frac{\sqrt{%
-\Delta }}{2a}\tanh \frac{\sqrt{-\Delta }}{2}\mu \left( x\right) +\frac{b}{2a%
}\right) ^{2}+\frac{\rho ^{2}}{\left( \frac{\sqrt{-\Delta }}{2a}\tanh \frac{%
\sqrt{-\Delta }}{2}\mu \left( x\right) +\frac{b}{2a}\right) ^{2}}  \notag \\
&&-2\lambda \sigma \left( \frac{\sqrt{-\Delta }}{2a}\tanh \frac{\sqrt{%
-\Delta }}{2}\mu \left( x\right) +\frac{b}{2a}\right) -\frac{2\rho \sigma }{%
\frac{\sqrt{-\Delta }}{2a}\tanh \frac{\sqrt{-\Delta }}{2}\mu \left( x\right)
+\frac{b}{2a}}  \notag \\
&&-\frac{\frac{\lambda \Delta }{4a}}{\cosh ^{2}\frac{\sqrt{-\Delta }}{2}\mu
\left( x\right) }+\frac{\frac{\rho \Delta }{4a}}{\cosh ^{2}\frac{\sqrt{%
-\Delta }}{2}\mu \left( x\right) \left( \frac{\sqrt{-\Delta }}{2a}\tanh 
\frac{\sqrt{-\Delta }}{2}\mu \left( x\right) +\frac{b}{2a}\right) ^{2}} 
\notag \\
&&+\sigma ^{2}+2\lambda \rho .  \TCItag{3.32}
\end{eqnarray}

As in the trigonometric case, the $\sigma =b=0$\ constraint leads to the
shape-invariant hyperbolic P\"{o}schl-Teller potential. Taking into account $%
\Delta =4ac<0,$\ the hyperbolic effective potential is reduced to%
\begin{equation}
V_{eff}^{\left( \text{Hyp.}\right) }\left( x,\mathbf{\lambda }\right) =\frac{%
\lambda c\left( \frac{\lambda }{a}-1\right) }{\cosh ^{2}\sqrt{-ac}\mu \left(
x\right) }-\frac{\rho a\left( \frac{\rho }{c}+1\right) }{\sinh ^{2}\sqrt{-ac}%
\mu \left( x\right) }-\left( \lambda \sqrt{\frac{c}{a}}-\rho \sqrt{\frac{a}{c%
}}\right) ^{2}.  \tag{3.33}
\end{equation}

\paragraph{Energy eigenvalues}

Substituting (3.8.c) in (3.5) we obtain the system of parametric equations%
\begin{eqnarray}
\lambda _{1}^{2}-\lambda _{0}^{2} &=&a\left( \lambda _{1}+\lambda
_{0}\right) ,  \TCItag{3.34.a} \\
2\left( \lambda _{1}\sigma _{1}-\lambda _{0}\sigma _{0}\right) &=&b\left(
\lambda _{1}+\lambda _{0}\right) ,  \TCItag{3.34.b} \\
\rho _{1}^{2}-\rho _{0}^{2} &=&-c\left( \rho _{1}-\rho _{0}\right) , 
\TCItag{3.34.c} \\
2\left( \rho _{1}\sigma _{1}-\rho _{0}\sigma _{0}\right) &=&-b\left( \rho
_{1}-\rho _{0}\right) ,  \TCItag{3.34.d} \\
q &=&c\left( \lambda _{1}+\lambda _{0}\right) -a\left( \rho _{1}-\rho
_{0}\right) \neq 0.  \TCItag{3.34.e}
\end{eqnarray}

From (3.34), we deduce the following recursion relations%
\begin{equation}
\rho _{k}=\left( -1\right) ^{k}\rho _{0}-\frac{c}{2}\left( 1-\left(
-1\right) ^{k}\right) ,\quad \lambda _{k}=\lambda _{0}+ak,\quad \sigma
_{k}=\sigma _{0}+bk.  \tag{3.35}
\end{equation}%
then the energy eigenvalues are given by%
\begin{eqnarray}
\mathcal{E}_{n} &=&\dsum\limits_{k=0}^{n-1}\mathcal{R}\left( \lambda
_{k},\sigma _{k},\rho _{k}\right)  \notag \\
&=&\dsum\limits_{k=0}^{n-1}\left( 4c\lambda _{k}+4a\rho _{k}+4\lambda
_{k}\rho _{k}-2b\sigma _{k}+\Delta \right)  \notag \\
&=&\left( a\rho _{0}+2\lambda _{0}\rho _{0}+c\lambda _{0}+\frac{ac}{2}%
\right) \left( 1-\left( -1\right) ^{n}\right) -n\left( 2a\rho _{0}+ac\right)
\left( -1\right) ^{n}  \notag \\
&&+n\left( 2c\lambda _{0}+ac-2b\sigma _{0}\right) +n^{2}\left(
ac-b^{2}\right)  \TCItag{3.36}
\end{eqnarray}

Taking into account the constraint $\sigma _{k}=b=0$\ as well as the sign of
the discriminant $\Delta \footnote{%
This means that $\func{sign}a=\func{sign}c$\ for $\Delta >0$\ and $\func{sign%
}a=-\func{sign}c$\ for $\Delta <0.$\ },$\ the energy spectra of the
bound-states for trigonometric and hyperbolic P\"{o}schl-Teller potentials
are%
\begin{equation}
\mathcal{E}_{n}^{\left( \text{Trig.}\right) }=\left[ \frac{1}{2}+\frac{%
\lambda _{0}}{a}+n-\left( a\rho _{0}+\frac{1}{2}\right) \left( -1\right) ^{n}%
\right] ^{2}-\left( \frac{\lambda _{0}}{a}-a\rho _{0}\right) ^{2}. 
\tag{3.37}
\end{equation}%
\begin{equation}
\mathcal{E}_{n}^{\left( \text{Hyp.}\right) }=-\left[ \frac{1}{2}+\frac{%
\lambda _{0}}{a}+n+\left( a\rho _{0}-\frac{1}{2}\right) \left( -1\right) ^{n}%
\right] ^{2}+\left( \frac{\lambda _{0}}{a}+a\rho _{0}\right) ^{2}. 
\tag{3.38}
\end{equation}

The energy eigenvalues thus obtained present, as for the case of
three-dimensional harmonic oscillator, some similar features with that
obtained in formulas (22) and (29) in Ref. [25], respectively.

\paragraph{Ground-state wavefunction}

The ground-state wavefunction $\psi _{0}\left( x\right) $\ associated to the
trigonometric and hyperbolic P\"{o}schl-Teller potentials (3.30) and (3.33)
are given respectively%
\begin{equation}
\psi _{0}^{\left( \text{Trig.}\right) }\left( x\right) =\frac{\mathcal{N}_{0}%
}{U\left( x\right) }\cos ^{\lambda /a}\sqrt{ac}\mu \left( x\right) \sin
^{-\rho /c}\sqrt{ac}\mu \left( x\right) .  \tag{3.39}
\end{equation}%
\begin{equation}
\psi _{0}^{\left( \text{Hyp.}\right) }\left( x\right) =\frac{\mathcal{N}_{0}%
}{U\left( x\right) }\cosh ^{\lambda /a}\sqrt{-ac}\mu \left( x\right) \sinh
^{\rho /c}\sqrt{-ac}\mu \left( x\right) .  \tag{3.40}
\end{equation}

\subsubsection{Three-dimensional Coulomb potential}

\paragraph{Superpotential and Effective potential}

Using (3.27.c), the function $\phi \left( x\right) $\ becomes%
\begin{equation}
\phi \left( x\right) =-\left[ \frac{2+b\mu \left( x\right) }{2a\mu \left(
x\right) }\right] ,  \tag{3.41}
\end{equation}%
and the corresponding superpotential and the effective potential read%
\begin{equation}
W\left( x,\mathbf{\lambda }\right) =-\lambda \left( \frac{2+b\mu \left(
x\right) }{2a\mu \left( x\right) }\right) -\frac{2a\rho \mu \left( x\right) 
}{2+b\mu \left( x\right) }+\sigma +\frac{U^{\prime }\left( x\right) }{2}, 
\tag{3.42}
\end{equation}%
\begin{eqnarray}
V_{eff}\left( x,\mathbf{\lambda }\right)  &=&\lambda ^{2}\left( \frac{2+b\mu
\left( x\right) }{2a\mu \left( x\right) }\right) ^{2}+\frac{4a^{2}\rho
^{2}\mu ^{2}\left( x\right) }{\left( 2+b\mu \left( x\right) \right) ^{2}}%
-\lambda \rho \left( \frac{2+b\mu \left( x\right) }{2a\mu \left( x\right) }%
\right)   \notag \\
&&-\frac{4a\rho \sigma \mu \left( x\right) }{2+b\mu \left( x\right) }-\frac{%
\lambda }{a\mu ^{2}\left( x\right) }+\frac{4a\rho }{\left( 2+b\mu \left(
x\right) \right) ^{2}}+2\rho \lambda +\sigma ^{2}.  \TCItag{3.43}
\end{eqnarray}

At first sight, the $\rho =0$\ constraint leads to the shape-invariant
three-dimensional Coulomb potential%
\begin{equation}
V_{eff}^{\left( \text{Cb.}\right) }\left( x,\mathbf{\lambda }\right) =\frac{%
\lambda }{a}\left( \frac{\lambda }{a}-1\right) \frac{1}{\mu ^{2}\left(
x\right) }+\frac{2\lambda }{a}\left( \frac{b\lambda }{2a}-\sigma \right) 
\frac{1}{\mu \left( x\right) }+\left( \frac{b\lambda }{2a}-\sigma \right)
^{2}.  \tag{3.44}
\end{equation}

\paragraph{Energy eigenvalues}

Since we are dealing with three-dimensional Coulomb potential, we will
assume that $\dfrac{\lambda }{a}\left( \dfrac{\lambda }{a}-1\right) =l\left(
l+1\right) $\ and $\dfrac{2\lambda }{a}\left( \dfrac{b\lambda }{2a}-\sigma
\right) =-Ze^{2},$\ it is straightforward to get%
\begin{equation}
\lambda =a\left( l+1\right) \ ,\qquad \sigma =\frac{b}{2}\left( l+1\right) +%
\frac{Ze^{2}}{2\left( l+1\right) },  \tag{3.45}
\end{equation}%
where $l$ is the angular momentum quantum number, $Z$\ the atomic number and 
$e$\ the electronic charge. Here the parameters $\lambda $\ and $\sigma $\
coincide with $\lambda _{0}$\ and $\sigma _{0},$\ respectively. As a
consequence to (3.45), the recursion relations (3.35) give%
\begin{equation}
\lambda _{k}=a\left( k+l+1\right) \ ,\qquad \sigma _{k}=\frac{b}{2}\left(
l+1\right) +\frac{Ze^{2}}{2\left( l+1\right) }+bk.  \tag{3.46}
\end{equation}%
for $k=0,1,2,\ldots $\ Since the potential (3.44) is obtained from the
restriction $\Delta =\rho _{k}=0,$ this requires that the coefficients $a,b$%
\ and $c$\ are nonzero and $\rho _{k}=$ $\left( -1\right) ^{k}\rho _{0}-%
\dfrac{c}{2}\left( 1-\left( -1\right) ^{k}\right) =0$ imposes the condition
that $k$\ should be even, i.e. $k=2p$ with $p\in 
\mathbb{N}
.$ As defined above in (3.36) and taking into account the last restriction,
the shape-invariance condition function $\mathcal{R}\left( \lambda
_{k},\sigma _{k},\rho _{k}\right) $\ satisfies%
\begin{eqnarray}
\mathcal{R}\left( \lambda _{kl},\sigma _{kl}\right)  &=&4c\lambda
_{kl}-2b\sigma _{kl}  \notag \\
&=&4ac\left( k+l+1\right) -2b\left[ \frac{b\left( l+1\right) }{2}+\frac{%
Ze^{2}}{2\left( l+1\right) }+bk\right]   \notag \\
&=&-\left( b^{2}k+\frac{Ze^{2}b}{l+1}\right) .  \TCItag{3.47}
\end{eqnarray}

The energy eigenvalues can now be obtained from equations (2.17) and (3.47).
By inserting $k=2p$\ in (3.47), the eigenvalues can be rewritten as%
\begin{eqnarray}
\mathcal{E}_{N,l}^{\left( \text{Cb.}\right) } &=&\dsum\limits_{k=0}^{N-1}%
\mathcal{R}\left( \lambda _{kl},\sigma _{kl}\right)   \notag \\
&=&-\dsum\limits_{p=0}^{\func{int}\left[ \frac{N-1}{2}\right] }\left(
2b^{2}p+\frac{Ze^{2}b}{l+1}\right)   \notag \\
&=&\frac{-\ b}{l+1}\left( 1+\func{int}\left[ \frac{N-1}{2}\right] \right)
\left( Ze^{2}+b\left( l+1\right) \func{int}\left[ \frac{N-1}{2}\right]
\right) .  \TCItag{3.48}
\end{eqnarray}%
where $\func{int}\left[ \dfrac{N-1}{2}\right] \ $is$\ $the greatest$\ $%
integer not larger than $\dfrac{N-1}{2}$.$\ $By definition, $\func{int}\left[
\dfrac{x}{2}\right] =\dfrac{x}{2}$\ for $x$\ even and $\dfrac{x-1}{2}$\ for $%
x$ odd. Here, both $l$\ and\ $N$\ are related by the relationship%
\begin{eqnarray}
\func{int}\left[ \dfrac{N-1}{2}\right]  &=&\frac{N-1-s}{2}  \notag \\
&=&\frac{2n-1-\left( 2l+1\right) }{2}  \notag \\
&=&n-l-1  \notag \\
&=&n_{r}.  \TCItag{3.49}
\end{eqnarray}%
with $s=2l+1$, $n$\ is the principal quantum number\ and $n_{r}$ a quantum
number which denotes the number of radial nodes for the wavefunction. Here,
we have assumed $N=2n$\ such that the number of bound-state levels is equal
to $\func{int}\left[ n+\dfrac{1}{2}\right] =n$ for $n\neq 0.$ However, we
can note, from (3.49), that the angular momentum quantum number $l$\
fulfills the condition $l\neq -\dfrac{1}{2},-1,-\dfrac{3}{2},\ldots $

If we make the replacement $b=\dfrac{Ze^{2}}{2\left( n_{r}+l+1\right) }+%
\dfrac{Ze^{2}}{2\left( l+1\right) },$\ then the energy eigenvalues (3.48)
become%
\begin{equation}
\mathcal{E}_{n_{r},l}^{\left( \text{Cb.}\right) }=\frac{-Z^{2}e^{4}\left(
1+n_{r}\right) }{4\left( n_{r}+l+1\right) ^{2}\left( l+1\right) ^{2}}\left(
n_{r}+2l+2\right) \left( n_{r}^{2}+2n_{r}+2\left( l+1\right) \left(
n_{r}+1\right) \right) .  \tag{3.50}
\end{equation}

For large values of $l$\ and $n_{r}$, we can make the approximation $l\sim
l+1$\ and $n_{r}\sim n_{r}+1$\ leading to impose that both quantum numbers
take all integral values from $0$\ to $l_{\max }$\ and $n_{r\max }$,
respectively, i.e.\ $l=0,1,2,\ldots ,l_{\max }$\ and $n_{r}=0,1,2,\ldots
,n_{r\max }$. Thus, the energy eigenvalues can now be obtained from (3.50) as%
\begin{eqnarray}
\mathcal{E}_{n_{r},l}^{\left( \text{Cb.}\right) } &=&-\ \frac{Z^{2}e^{4}}{%
4\left( n_{r}+l+1\right) ^{2}\left( l+1\right) ^{2}}n_{r}^{2}\left(
n_{r}+2l+2\right) ^{2}  \notag \\
&=&-\ \left[ \frac{Ze^{2}-\kappa F\left( n_{r},l\right) }{2\ \left(
n_{r}+l+1\right) }\right] ^{2},  \TCItag{3.51}
\end{eqnarray}%
with $F\left( n_{r},l\right) =n_{r}^{2}+\left( l+1\right) \left(
2n_{r}+1\right) $\ and $\kappa =\dfrac{Ze^{2}}{l+1}.$\ From general
considerations, it is evident that the spectrum of negative eigenvalues of
the energy will be discrete, while that of the positive eigenvalues will be
continuous. It follows from (3.51) that the function $F\left( n_{r},l\right) 
$\ is fulfilled by the condition%
\begin{equation}
F\left( n_{r},l\right) =n_{r}^{2}+\left( l+1\right) \left( 2n_{r}+1\right) <%
\frac{Ze^{2}}{\kappa }.  \tag{3.52}
\end{equation}

The energy spectra given in (3.51) have already been established by Quesne
and Tkachuk [18,19].

\paragraph{Ground-state wavefunction}

Let us now complete this algebraic determination of energy eigenvalues by a
construction of the corresponding \textquotedblright
radial\textquotedblright\ ground-state wavefunction $\psi _{0l}\left(
x\right) $. Using the last considerations, the superpotential can be
rewritten as $W_{l}\left( x\right) =-\dfrac{l+1}{\mu \left( x\right) }+%
\dfrac{Ze^{2}}{2\left( l+1\right) }+\dfrac{U^{\prime }\left( x\right) }{2}$
and the \textquotedblright radial\textquotedblright\ ground-state
wavefunction is given through (2.20) by%
\begin{equation}
\psi _{0l}\left( x\right) =\frac{\mathcal{N}_{0}}{U\left( x\right) }\mu
^{l+1}\left( x\right) \exp \left[ -\frac{Ze^{2}}{2\ \left( l+1\right) }\mu
\left( x\right) \right] .  \tag{3.53}
\end{equation}

\section{Conclusion}

In the present paper, we have generated four \textquotedblright
new\textquotedblright\ solvable, real and shape-invariant potentials simply
by judicious applications of the supersymmetric quantum mechanics formalism
and shape-invariant potentials. In all cases, we have derived the effective
potentials as well as their accompanying energy spectra of bound-states and
ground-state wavefunctions. However, the new and the important contribution
of this paper is to point out how the simple fact of going from the positive
case $\left( \pi =1\right) $,\ characterized by an equidistant spectra, to
the negative one $\left( \pi =-1\right) $, allows to generate a new
shape-invariant three-dimensional harmonic oscillator, Morse and P\"{o}%
schl-Teller (I and II) potentials with non-equidistant spectra (3.16),
(3.24) and (3.36), respectively, while the three-dimensional Coulomb
potential has only a finite number of bound-states (3.51) in contrast with
the standard coulomb problem.

To conclude, the supersymmetric quantum mechanics can also be a useful
machinery for the treatment of wide classes of potentials. It may be
possible to explore, in the context of position-dependent effective mass
within the framework of non compact $\mathbf{SO}\left( 2,2\right) $\ Lie
algebra, all shape-invariant potentials listed and tabulated in Refs.
[19,21], respectively. The works are in progress and will be deferred to
later publication.

\end{document}